\documentclass[a4paper]{jpconf}
\usepackage{graphicx}
\begin{document}
\title{Underground tests of quantum mechanics. \\Whispers in the cosmic silence?}

\author{C~Curceanu$^1$$^{,2,}$$^{9}$, S~Bartalucci$^1$, A~Bassi$^3$, M~Bazzi$^1$, S~Bertolucci$^4$, C~Berucci$^1$$^{,5}$, A~M~Bragadireanu$^1$$^{,2}$, M~Cargnelli$^5$, A~Clozza$^1$, L~De~Paolis$^1$, S~Di~Matteo$^6$, S~Donadi$^3$, J-P~Egger$^7$, C~Guaraldo$^1$, M~Iliescu$^1$, M~Laubenstein$^8$, J~Marton$^5$, E~Milotti$^3$, A~Pichler$^5$, D~Pietreanu$^1$$^{,2}$, K~Piscicchia$^1$$^{,9}$, A~Scordo$^1$, H~Shi$^5$, D~Sirghi$^1$$^{,2}$, F~Sirghi$^1$$^{,2}$, L~Sperandio$^1$, O~Vazquez Doce$^{10}$, J~Zmeskal$^5$}


\address{$^1$INFN, Laboratori Nazionali di Frascati, CP 13, Via E. Fermi 40, I-00044, Frascati (Roma), Italy}

\address{$^2$``Horia Hulubei''National Institute of Physics and Nuclear Engineering, Str. Atomistilor no. 407, P.O. Box MG-6,  Bucharest - Magurele, Romania}

\address{$^3$Dipartimento di Fisica, Universit\`{a} di Trieste and INFN-- Sezione di Trieste, Via Valerio, 2, I-34127 Trieste, Italy}

\address{$^4$University and INFN Bologna, Via Irnerio 46, I-40126, Bologna, Italy}

\address{$^5$The Stefan Meyer Institute for Subatomic Physics,  Boltzmanngasse 3, A-1090 Vienna, Austria}

\address{$^6$Institut de Physique UMR CNRS-UR1 6251, Universit\'e de Rennes1, F-35042 Rennes, France}

\address{$^7$Institut de Physique, Universit\'e de Neuch\^atel, 1 rue A.-L. Breguet, CH-2000 Neuch\^atel, Switzerland}

\address{$^8$INFN, Laboratori Nazionali del Gran Sasso, S.S. 17/bis, I-67010 Assergi (AQ), Italy}

\address{$^9$Museo Storico della Fisica e Centro Studi e Ricerche ``Enrico Fermi'', Roma, Italy}

\address{$^{10}$Excellence Cluster Universe, Technische Universit\"at M\"unchen, Garching, Germany}

\ead{catalina.curceanu@lnf.infn.it}

\begin{abstract}
By performing X-rays measurements in the ``cosmic silence''  of the underground laboratory of Gran Sasso, LNGS-INFN, we test a basic principle of quantum mechanics: the Pauli Exclusion Principle (PEP), for electrons. We present the achieved results of the VIP experiment and the ongoing VIP2 measurement aiming to gain two orders of magnitude improvement in testing PEP. We also use  a similar experimental technique to
 search for radiation (X and gamma) predicted by continuous spontaneous localization models, which aim to solve the ``measurement problem''. 
\end{abstract}

\section{Introduction}
To the best of our knowledge, there are 2 spin-separated classes of particles: fermions, with half-integer spin, and bosons, with integer spin. The Pauli Exclusion Principle (PEP), only valid for fermions, states that two fermions can not be in the same quantum state \cite{Pauli}. PEP is a fundamental principle in physics, for which an intuitive explanation is still missing \cite{Feynman}. The violation of PEP would certainly be related to new physics, beyond the Standard Model. A thorough test of the PEP is than necessary for all types of particles.

In the past, very few experiments have tested the PEP. Some of them put stringent limits to the probability of the violation of the Pauli Exclusion Principle in fermionic systems, for example the experiment performed by the DAMA collaboration \cite{DAMA}, investigating transitions of stable systems from a non-Pauli violating state to a Pauli violating state. These transitions violate the Messiah-Greenberg superselection rule \cite{MG}, which forbids the change of the symmetry  of the state of a stable system. To bypass this rule, the VIP experiment introduces ``new'' electrons in the system by means of an electric current. 


In the next sections we describe the experimental method, the VIP setup and the  results obtained by running  VIP at the underground
 Gran Sasso National Laboratory (LNGS). We present also the VIP2 setup, a major upgrade of VIP, which is using fast
 Silicon Drift Detectors (SDD) and a veto system, aiming to gain two orders of magnitude in PEP violation study.

We will then conclude the paper by presenting some ideas to use similar experimental techniques
to perform measurements of X-rays spontaneously emitted by electrons and protons in matter, predicted by the collapse models \cite{catan1,catan2,catan3,catan4}.
The collapse models deal with the ``measurement problem" in quantum mechanics,
by introducing a new physical dynamics that naturally collapses the state
vector and make predictions which differ from those of standard quantum
mechanics \cite{adler}. In this context, one of the most exciting tasks is to perform cutting-edge
experiments, in order to asses whether quantum mechanics is exact, or an
approximation of a deeper level theory.

\section{The experimental method and VIP results}
In the VIP and VIP2 experiments ``new'' electrons are introduced into a copper target by circulating an electric current.
The VIP setup consisted of a copper cylindrical target, 45 mm radius, 50 $\mu$m thickness, 
and 88 mm height, where a current of 40 A was circulated, surrounded by 16 equally spaced Charge Coupled Devices (CCDs).
The CCDs were at a distance of 23 mm from the copper cylinder, and paired
one above the other.
For the VIP2 setup, the cooper target consists of 2 strips with 25 $\mu$m thickness and 6 cm length, where an electric current of 100 A is circulated.

\begin{figure}[h]
 \centering
 \includegraphics[height=1.1in]{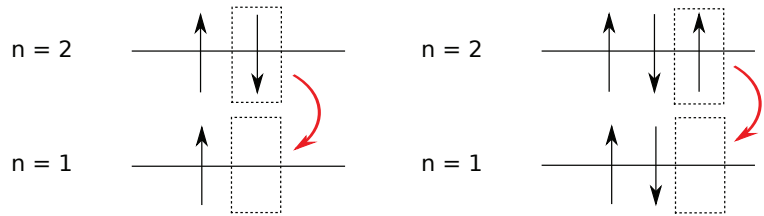}
 \caption{Normal 2p to 1s transition with an energy  around 8 keV for Copper (left) and Pauli-violating 2p to 1s transition with a transition energy around 7,7 keV in Copper (right).}
 \label{fig:energy_scheme}
\end{figure}

 The ``new'' electrons  introduced into the copper target have a certain probability to interact with and be captured by copper atoms. In the course of this interaction, the electrons might form a new symmetry state with the electrons of the atom. This process is the reason why the VIP experiment does not violate the Messiah-Greenberg superselection rule, since the current electrons are ``new'' to the electrons in the atom, i.e. they have no predefined symmetry  of the state. With a certain probability which should be determined by the experiment, the newly formed symmetry state might have a symmetric component and the electron could undergo transitions from the 2p level to the 1s one, with the 1s already occupied by 2 electrons, which transition obviously violates the PEP, Figure \ref{fig:energy_scheme}.

 The non-Paulian transition has about 300 eV lower energy with respect to the allowed one. The energy of the X-rays which are emitted during the cascading process are recorded by X-ray detectors, which are mounted close to the copper target. The recorded spectrum is then analyzed looking for an excess of counts over the background in the energy range of the forbidden transition, which is around 7,7 keV. The excess of counts or the lack of them will either signal discovery of the violation of the PEP or, else, determine a new upper bound for its violation.
%

The VIP experiment was installed at LNGS-INFN in Spring 2006 and was taking data until 2010, alternating period with current on
(signal) to periods with current off (background). The data analysis resulted in an upper limit for the violation of the PEP of $4.7 \times 10^{-29}$ \cite{catan1, catan3}. 


\section{The VIP2 experiment}

In the VIP2 experiment the CCDs were replaced by Silicon Drift Detectors (SDD), which have a superior energy resolution and offer timing capability. Six SDDs units, with a total active area of 6 cm$^2$ are mounted close to the Cu target, giving an acceptance which is about ten times as large as the acceptance of VIP CCDs. Moreover, an active shielding system (veto) was implemented, to reduce the background in the energy region of the forbidden transition. These systems will have an important contribution to improve the limit for the violation of the PEP by two orders of magnitude with new data which are presently comming by running the VIP2 experiment at LNGS \cite{Mar13}.

\begin{figure}[htbp]
  \centering
\includegraphics[width=15cm,clip]{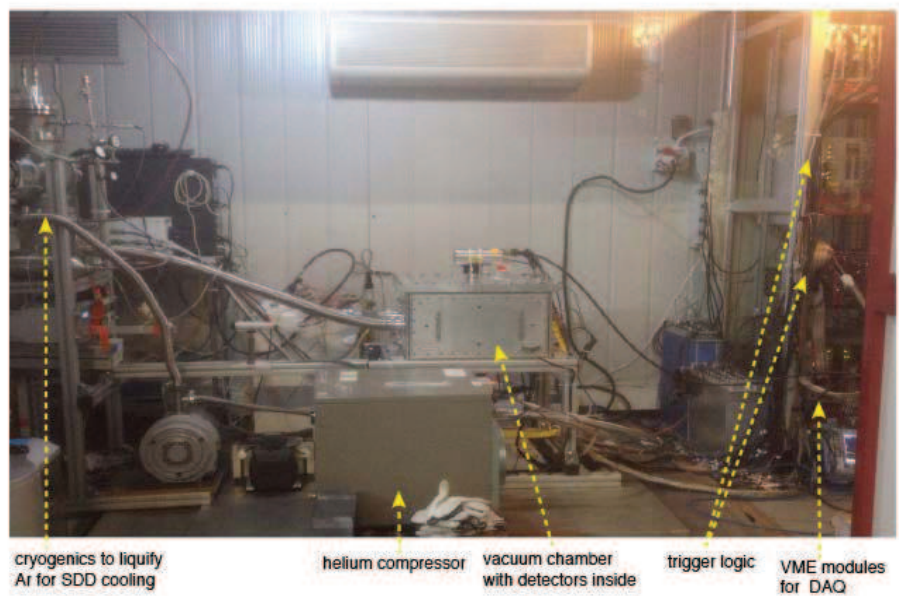}
\vspace{-0.5cm}
\caption{
         A picture of the VIP-2 setup in the barrack at LNGS . 
         }
\label{fig:barrack}       
\end{figure}

In November 2015, the VIP2 setup was installed at Gran Sasso  \cite{kaku}.
In the year 2016, data with the complete VIP2 detector system at the LNGS without shielding were taken.
In Figure \ref{fig:barrack} the VIP2 setup as installed at LNGS is shown.

\begin{figure}[htbp]
\centering
\includegraphics[width=13cm,clip]{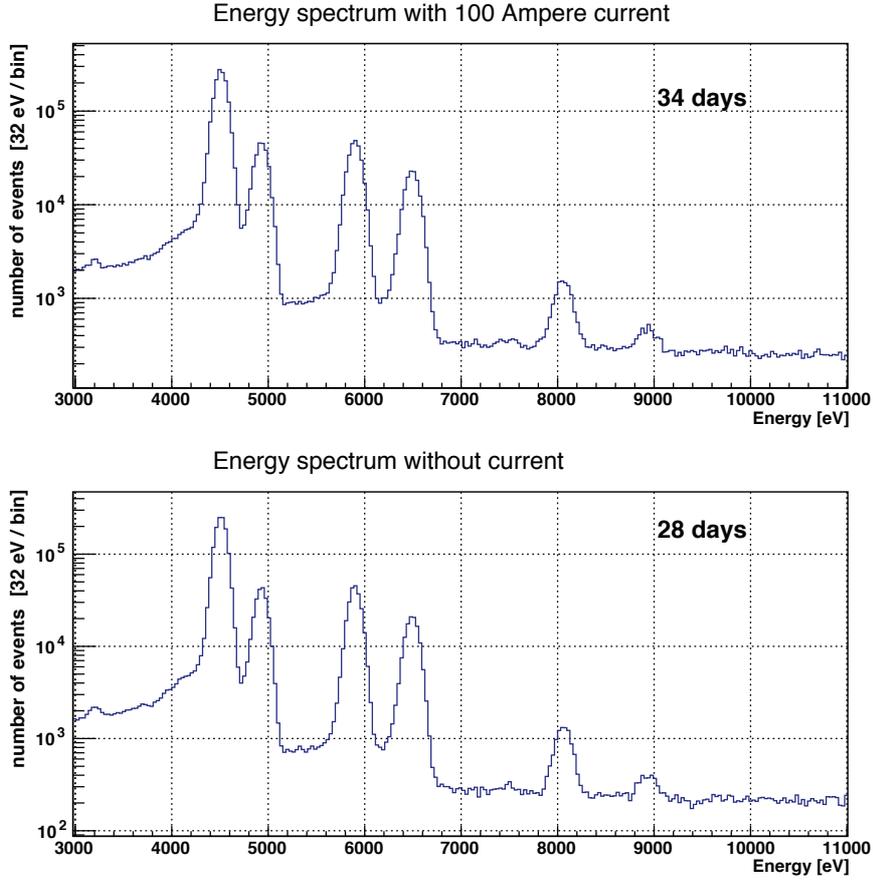}
\caption{
         Upper figure: the energy spectrum obtained by VIP2 in 34 days with current (I=100~A); Lower figure: the energy spectrum obtained by VIP2 in 28 days without current. 
         }
\label{fig:kaku}       
\end{figure}

Data with 100 Ampere DC current applied to the copper strip for 34 days was collected.
Together with the data collected for 28 days without current,
an analysis of the two spectra to determine a new value of the upper limit for PEP violation is on going.
Figure \ref{fig:kaku} shows the summation of the data from 2016 of all the SDDs; the preliminary analysis shows the energy resolution of the summed spectra at 8 keV is less than 190 eV FWHM.

In table 1 \cite{Mar13} the improvement factors of VIP2, with respect to VIP, are presented.
 In the coming 3 years of run we will gain two orders of magnitude in the PEP violation probability or will have signals of a violation.

\begin{table}
\caption{The improvement factors for VIP2 in comparison to VIP (in parenthesis).}  
\begin{center}
{\begin{tabular}[b]{l l l}
\hline
\hline
Changes in VIP2     &  value VIP2 (VIP)  &    expected gain  \\
\hline
acceptance          & 12 \% ($\sim$ 1 \%)     &  12  \\
increase current    & 100 A (40 A)    & $>$ 2   \\
reduced length      & 3 cm (8.8 cm )  &  1/3 \\
\hline
total linear factor &              &     8  \\
\hline 
energy resolution &  170 eV (320 eV) @ 8 keV   &  4 \\   
reduced active area  & 6 cm $^2$ (114 cm $^2$)   &   20 \\
better shielding and veto   &         &    5-10 \\
higher SDD efficiency   &   &    1/2 \\
\hline 
background reduction  &  &  200 - 400 \\
\hline 
overall improvement  &  &   $>$ 120 \\
\hline
\end{tabular}}
\end{center}
\end{table}

\section{X-ray measurements for testing the dynamical reduction models}

The aim of the Dynamical Reduction Models (DRM) is to solve the  so-called ``measurement problem'' in Quantum Mechanics (QM). The linear and unitary nature of the Shr\"odinger equation allows, in principle, the superposition of macroscopic states, but such superpositions are not observed in the measurement process, which is intrinsically non-linear and stochastic \cite{neumann,bassi}. The measurement problem led to the introduction of the wave packet reduction principle which, nevertheless, does not predict the scale at which the quantum-to-classical transition occurs, nor explains the collapse mechanism.    

The work of Ghirardi, Rimini and Weber \cite{ghi} lead to the development of a consistent DRM known as Quantum Mechanics with Spontaneous Localization (QMSL). According to the QMSL model
each particle of a macroscopic system of $n$ distinguishable particles experiences sudden spontaneous localizations, on the position basis, with a mean rate $\lambda =10^{-16}$ s$^{-1}$, and a correlation length $a=10^{-7}$ m. Between two localizations particles evolve according to the Shr\"odinger dynamics. The model ensures, for the macroscopic object, the decoupling of the interanl and Center of Mass (CM) motions. The interanl motion is not affected by the localization, whereas the CM motion is localized with a rate $\lambda_{macro}= \, n \, \lambda$.

Subsequently, the theory was developed in the language of the non-linear and stochastic Shr\"odinger equation \cite{pear,ghi-pe}, where besides the standard quantum Hamiltonian, two other terms induce a diffusion process for the state vector, which causes the collapse of the wave function in space. In its final version \cite{pe-sq} the model is known as the mass proportional Continuous Spontaneous Localization (CSL).

The value of the mean collapse rate is presently argument of debate. According to CSL $\lambda$ should be of the order of $10^{-17}$ s$^{-1}$, whereas a much stronger value $10^{-8\pm 2}$ s$^{-1}$ was proposed by S. L. Adler \cite{adler1} based on arguments related to the latent image formation and the perception of the eye.
 
DRM posses the unique characteristic to be experimentally testable, by measuring the (small) predicted deviations with respect to the standard quantum mechanics. The conventional approach is to generate spatial superpositions of mesoscopic systems and examine the loss of interference, while environmental noises are, as much as possible, under control. The present day technology, however, does not allow to set stringent limits on $\lambda$ by applying this method. The most promising testing ground, instead, is represented by the search for the spontaneous radiation emitted by charged particles when interacting with the collapsing stochastic field \cite{fu}. A measurement of the emitted radiation rate thus enables to set a limit on the $\lambda$ parameter of the models.

The radiation spectrum spontaneously emitted by a free electron, as a consequence of the interaction with the stochastic field, was calculated by Q. Fu \cite{fu} in the framework of the non-relativistic CSL model, and it is given by:

\begin{equation}\label{furate}
\frac{d\Gamma (E)}{dE} = \frac{e^2 \lambda}{4\pi^2 a^2 m^2 E}
\end{equation}
in eq. (\ref{furate}) $m$ represents the electron mass and $E$ is the energy of the emitted photon. In the mass proportional CSL model the stochastic field is assumed to be coupled to the particle mass density, then the rate is to be multiplied by the factor $(m/m_N)^2$, with $m_N$ the nucleon mass. Using the measured radiation appearing in an isolated slab of Germanium \cite{miley} corresponding to an energy of 11 KeV, and employing the predicted rate eqn. (\ref{furate}), Fu obtained the following upper limit for $\lambda$ (non-mass poportional model):

\begin{equation}\label{fulimit2}
\lambda < 0.55 \cdot 10^{-16} s^{-1}.
\end{equation}
In eq. (\ref{fulimit2}) the QMSL value for $a$ ($a=10^{-7}$ m) is assumed and the four valence electrons were considered to contribute to the measured X-ray emission, since the binding energy is $\sim 10$ eV in this case, and they can be considered as \emph{quasi-free}. Recent re-analyses of Fu's work \cite{adler1,mullin} corrected the limit in Eq. \ref{fulimit} to $\lambda < 2 \cdot 10^{-16} s^{-1}$.

We improved the limit on the collapse rate \cite{catakristian} by analysing the data collected by the IGEX (International Germanium EXperiment) experiment \cite{igex1}.
IGEX is a low-background experiment based on low-activity Germanium detectors dedicated to the $\beta \beta 0 \nu$ decay research.
We performed a fit of the published X-ray emission spectrum \cite{igex2}, which refers to an 80 kg day exposure, in the energy range $\Delta E =$ $4.5\div 48.5$ KeV $\ll m$. The energy interval is compatible with the non-relativistic assumption of the model (Eq. (\ref{furate})).

A Bayesian model was adopted to calculate the $\chi^2$ variable minimized to fit the X ray spectrum, assuming the predicted (Eq. (\ref{furate})) energy dependence:

\begin{equation}\label{fitfunc}
\frac{d\Gamma (E)}{dE} = \frac{\alpha(\lambda)}{E}.
\end{equation}
The obtained values for $\lambda$ are: 

\begin{equation}\label{fulimit}
\lambda \leq 2.5 \cdot 10^{-18} s^{-1},
\end{equation}
if no mass dependence is considered, and 

\begin{equation}\label{no mass}
\lambda \leq 8.5 \cdot 10^{-12} s^{-1},
\end{equation}
in the mass proportional CSL assumption. This analysis improves the previous limit \cite{fu} of two orders of magnitude.

By using a similar method, we are presently performing a dedicated experiment at LNGS which will allow for 1 - 2  orders of magnitude further improvement on the collapse rate parameter $\lambda$.

\section{Conclusions}

We presented the VIP experiment searching for the ``impossible atoms'', i.e., atoms where the Pauli Exclusion Principle (PEP) is violated. We have set a limit on PEP violation probability for electrons as $\frac{\beta^{2}}{2} < 4.7 \times 10^{-29}$, and we will improve this limit by two orders of magnitude, or find the PEP violation, in the coming 3 years by running the VIP2 setup at the LNGS underground laboratory. By using a similar experimental technique we are exploring the collapse theories, looking for spontaneously emitted radiation, to set stringent limits on the $\lambda$ parameter characterizing these models.
X-rays measurements performed in the ``cosmic silence'' of the LNGS-INFN laboratory, are a very important tool to test quantum mechanics.

\ack

We thank H. Schneider, L. Stohwasser, and D. St\"{u}ckler from Stefan-Meyer-Institut 
for their fundamental contribution in designing and building the VIP2 setup. 

We acknowledge the very important assistance of the INFN-LNGS laboratory staff during all phases of preparation, 
installation and data taking.

The support from the EU COST Action CA 15220 is gratefully acknowledged.
 
We thank the Austrian Science Foundation (FWF) which supports the VIP2 project with the grant P25529-N20 and  Centro Fermi for the grant  ``Problemi aperti nella meccania quantistica''.

Furthermore, this paper was made possible through the support of a grant from the Foundational Questions Institute, FOXi (``Events'' as we see them: experimental test of the collapse models as a solution of the measurement problem) and a grant from the John Templeton Foundation (ID 581589).
The opinions expressed in this publication are those of the authors and do not necessarily reflect the views of the John Templeton Foundation.

\smallskip

\end{document}